\documentclass[letterpaper]{article} 
\usepackage{aaai25}  
\usepackage{times}  
\usepackage{helvet}  
\usepackage{courier}  
\usepackage[hyphens]{url}  
\usepackage{graphicx} 
\usepackage{natbib}  
\usepackage{caption} 
\usepackage{algorithm}
\usepackage{algorithmic}

\usepackage{amsmath}
\usepackage{amssymb}
\usepackage{booktabs}
\usepackage{epsfig}
\usepackage{enumerate}
\usepackage[inkscapelatex=false]{svg}
\usepackage{newfloat}
\usepackage{listings}
\DeclareCaptionStyle{ruled}{labelfont=normalfont,labelsep=colon,strut=off} 
\floatstyle{ruled}
\newfloat{listing}{tb}{lst}{}
\floatname{listing}{Listing}
\title{Practicable Black-box Evasion Attacks on Link Prediction in Dynamic Graphs\\ ---A Graph Sequential Embedding Method}
\author {
    Jiate Li\textsuperscript{\rm 1 \rm 2},
    Meng Pang\textsuperscript{\rm 1}$^*$,
    Binghui Wang\textsuperscript{\rm 2}$^*$
}
\affiliations {
    \textsuperscript{\rm 1}Nanchang University, Nanchang, China\\
        \textsuperscript{\rm 2}Illinois Institute of Technology, Chicago, USA\\
    \textsuperscript{\rm *}Corresponding authors\\

    jetrichardlee@gmail.com, pangmeng1992@gmail.com,
    bwang70@iit.edu
}

\usepackage{bibentry}

\begin{document}

\maketitle

\begin{abstract}
Link prediction in dynamic graphs (LPDG) 
has been widely applied to real-world applications such as website recommendation, traffic flow prediction, organizational studies, etc. These models are usually kept local and secure, with only the interactive interface restrictively available to the public. Thus, the problem of the black-box evasion attack on the LPDG model, where model interactions and data perturbations are restricted, seems to be essential and meaningful in practice. In this paper, we propose the first practicable black-box evasion attack method that achieves effective attacks against the target LPDG model, within a limited amount of interactions and perturbations. To perform effective attacks under limited perturbations, we develop a graph sequential embedding model to find the desired state embedding of the dynamic graph sequences, under a deep reinforcement learning framework. To overcome the scarcity of interactions, we design a multi-environment training pipeline and train our agent for multiple instances, by sharing an aggregate interaction buffer. Finally, we evaluate our attack against three advanced LPDG models  on three real-world graph datasets of different scales and compare its performance with related methods under the interaction and perturbation constraints. Experimental results show that our attack is both effective and practicable.
\end{abstract}


\begin{links}
    \link{Code}{https://github.com/JetRichardLee/GSE-METP}
\end{links}

\section{Introduction}
\label{sec:intro}

\begin{figure}[htbp]
  \centering
   \includegraphics[width=\linewidth]{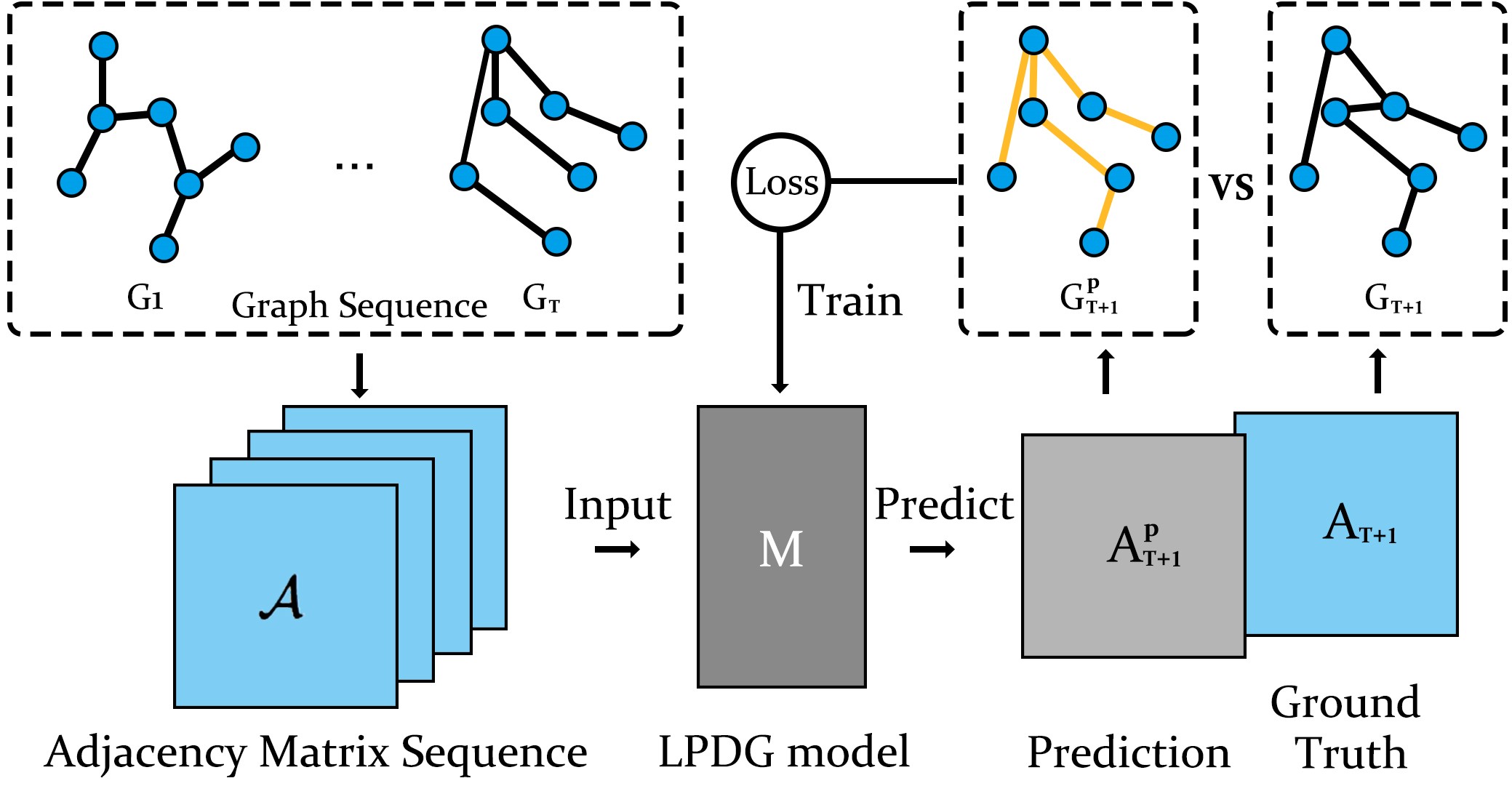}
   \caption{LPDG: a dynamic graph sequence $\mathcal{G} = \{G_{t}, t \in [1,T]\}$ is taken as an input, usually represented by an adjacency matrix sequence.  LPDG  predicts the future graph in the next time slice $G_{T+1}^p$. It is trained to increase the accuracy metric compared with the ground truth $G_{T+1}$.}
   \label{fig:LPDG}
\end{figure}

Dynamic graph sequences are a crucial structure for capturing the temporal dynamics of real-world systems. They represent the timing information within a dynamic graph, such as the historical evolution of interactions in a system. Link Prediction in Dynamic Graphs (LPDG), illustrated in Figure \ref{fig:LPDG}, leverages a dynamic graph sequence as input to predict future links for the next timestamp. Accurate predictions are highly valuable for a wide range of applications, including online recommendations, traffic management, and modeling disease contagion.
In recent years, numerous methods for LPDG have been developed, including DynGEM~\cite{DBLP:journals/corr/abs-1805-11273}, MTCP~\cite{MTCP}, EvolveGCN~\cite{pareja2020evolvegcn}, DyGCN~\cite{dygcn}, and MetaDyGNN~\cite{10.1145/3488560.3498417}. These methods differ in their prediction settings. For example, MTCP and DyGCN focus on predicting the entire future graph at the next timestamp, whereas EvolveGCN and DynGEM provide predictions for the next set of changing edges.


Like adversarial attacks on static graphs, recent work~\cite{fan2021reinforcement} shows that LPDG for dynamic graphs is also vulnerable to \emph{black-box} evasion attack, a practical attack 
where the attacker has 
no access to parameters or structure of the target model.
However, the existing attack (called SAC)~\cite{fan2021reinforcement} based on reinforcement learning (RL) 
requires millions of interactions with the target model. Despite that, due to a coarse embedding method on the graph sequence, SAC is unable to attack large-scale graphs with large state spaces. These limitations  
inhibit SAC from being a practicable attack method in real-world scenarios.

We propose the \emph{first practicable} black-box evasion attack on  LPDG methods. Given a target model and instance, we generate the perturbed instance 
with 
only a few interactions with the model, while achieving  promising attack results. Our attack method mainly consists of two novel designs:
\begin{enumerate}[1)]

\item \textbf{A graph sequential embedding (GSE)}. GSE has a static degree embedding for every graph in the sequence, which enables it to find the desired state representations of the embedding sequence using two Long Short-term Memory (LSTM)~\cite{LSTM} networks. Later, these representations can be applied to perform the black-box attack based on a deep deterministic policy gradient (DDPG)~\cite{pmlr-v32-silver14} framework.



\item \textbf{A multi-environment training pipeline (METP)}. Considering that our model does not have sufficient resources for RL due to the limit on the interactions per instance, we  hypothesize that there are similarities between instances from the same dataset and the same target model, and that training experience can be shared between them. Based on this hypothesis, we design a multi-environment training pipeline that sets multiple target instances as environments and trains a single model to attack them all.

     
     
\end{enumerate} 

Via testing on  {three} recent LPDG models,  
Dynamic Graph Convolutional Network (DyGCN)~\cite{dygcn}, Attention Based Spatial-Temporal Graph Convolutional Network (ASTGCN)~\cite{Guo_Lin_Feng_Song_Wan_2019} and  {Hyperbolic Temporal Graph Network (HTGN)~\cite{9999499}}, 
over three real-world datasets, our method proves to be both reliable and effective under the practicable constraints and achieves state-of-the-art performance. Ablation experiments demonstrate the rationality of our graph sequential embedding design and multi-environment design.

Our main contributions are summarized as follows:
\begin{itemize}
\item[$\bullet$] We propose the first practicable black-box evasion attack against LPDG, 
which learns effective attacks by only a few amount of interactions with the target model.

\item[$\bullet$] We develop a GSE method that assists to find reliable state embeddings for the RL framework to learn effective attacks. Furthermore, we design a METP  
to overcome the constraint of target model interactions by sharing experience between multiple instances.

\item[$\bullet$] 
Evaluation results demonstrate the superiority of our 
practicable black-box attacks. Ablation experiments prove the rationality and effectiveness of our designs.
\end{itemize}
\section{Related Work}
\subsection{Adversarial Attacks on Static Graphs}

Existing attacks on graph learning for static graphs are classified as 
poisoning attacks and 
{evasion attacks}.
Poisoning attack \cite{dai2018adversarial,zugner2019adversarial,xu2019topology,takahashi2019indirect,liu2019unified,sun2020adversarial,zhang2020backdoor,wang2023turning,yang2024distributed} is performed in the training phase (and testing phase). 
In training-time poisoning attacks, given a graph learning algorithm and a graph, an attacker carefully perturbs the graph (e.g., inject new edges to or remove the existing edges
from the graph, perturb the node features) in the training phase, such that the learnt model misclassifies as many  nodes, links, or graphs as possible on the clean testing graph(s). A special type of poisoning attack, called backdoor attack \cite{zhang2020backdoor}, also perturbs testing graphs in the testing phase. Further,     
\citet{yang2024distributed} generalize the backdoor attack on graphs in the federated learning setting \cite{wang2022graphfl}. 

Evasion attack \cite{dai2018adversarial,xu2019topology,wang2019attacking,wu2019adversarial,ma2019attacking,ma2020towards,li2021adversarial,mu2021hard,wang2022bandits,wang2023turning,wang2024efficient} is performed in the inference phase. In these attacks, 
given a graph learning model and clean graph(s), an attacker carefully perturbs the graph structure or/and node features such that as many nodes or graphs as possible are misclassified on the perturbed graph(s) by the given model. 


\subsection{Deep Deterministic Policy Gradient}
RL approaches with the actor-critic structure have been proposed and received great attention. In~\citet{haarnoja2018soft}, the soft actor-critic (SAC) is described, with an additional reward on the entropy of the policy distribution. In~\citet{schulman2017proximal}, it proposes the proximal policy optimization (PPO), based on the trust region policy optimization (TRPO)~\cite{schulman2015trust}, with the clipped surrogate objective added to the loss function as refinement. In~\cite{pmlr-v32-silver14}, the deterministic policy gradient (DPG), whose policy gives a deterministic action instead of a possibility distribution, is proposed. And in~\cite{lillicrap2015continuous}, deep deterministic policy gradient (DDPG) is proposed with the extension of deep reinforcement learning to DPG.
\section{Practicable Black-box Evasion Attack}
In this section, we first define
the practicable black-box evasion attack problem, and then introduce our two main designs, the graph sequential embedding  (GSE) and the multi-environment training pipeline (METP).  GSE learns the hidden sequential features of the dynamic graph sequence, and gives the embedding states, which are used by the policy network and the Q network of a DDPG agent. The agent's policy network and Q network are trained under the METP and propagate gradients back to train the GSE models. 

\subsection{Problem Definition}
\subsubsection{LPDG Model}
Simulating a real-world attacking scenario, the LPDG model is private on its structure and parameters, but public on prediction interactions in limited chances. We only have the interface for inputting a dynamic graph sequence and receiving the prediction.
\paragraph{Clean input data} A finite sequence of undirected graphs $\mathcal{G} = \{ G_{1},G_{2},...,G_{T}\}$  is given as the input data, where $G_{t} = (V_{t},E_{t}),\forall t\in [1:T] $ denotes a snapshot graph at the time slice $t$. $V_{t}$ is the set of nodes and $E_{t}$ is the set of edges. In our attack setting, we could perform limited perturbations on the edge data $\mathcal{E} = \{E_{t},\forall t\in [1:T] \}$.

\paragraph{Output and metric}  The target LPDG model $M$ is a black-box function. It predicts the edges of the graph at the next time slice $G_{T+1}^{p} = M(\mathcal{G}) = (V_{T+1}^{p},E_{T+1}^{p})$ and evaluates the performance by comparing $E_{T+1}^{p}$ with the ground truth $E_{T+1} \in G_{T+1}$. We use the F1 score as a metric to evaluate the prediction and expect the attack to fool the model by having as low F1 
as possible. 
\begin{equation}\small
\begin{aligned}
& Precision = \frac{|E_{T+1}\cap E_{T+1}^{p}|}{|E_{T+1}^{p}|}\\
& Recall = \frac{|E_{T+1}\cap E_{T+1}^{p}|}{|E_{T+1}\cap E_{T+1}^{p}|+|E_{T+1}\cap \complement E_{T+1}^{p}|}\\
& F_{1} = \frac{2*Precision*Recall}{Precision+Recall}
\end{aligned}
\end{equation}

\subsubsection{Attack Under Reinforcement Learning}
\label{sec:3.1.2}
Since we have no knowledge of the LPDG model but only chances to interact with it, it requires us to learn from experience generated by limited exploration. Therefore, we apply a reinforcement learning method as our basic framework to perform the attack. We define our attack problem as follows.

\paragraph{State} \label{state} The perturbed dynamic graph sequence $\hat{\mathcal{G}} = \{ \hat{G_{1}},\hat{G_{2}},...,\hat{G_{T}}$ \} describes the ground information during the attack. Since we only care about the $\hat{E}^p_{T+1} \in \hat{G}^{p}_{T+1}$ in the link prediction result, we assume that the nodes in the graph sequence are fixed as $\hat{V}_{t} = V, t\in[1,T+1]$ and basically, use an adjacency matrix sequence $\hat{\mathcal{A}}$ = \{ $\hat{A}_{1},\hat{A}_{2},...,\hat{A}_{T}$ \} to represent the state of the dynamic graph sequence. This raw representation has a large state space and lacks the essential feature extraction, making the policy and the Q difficult to converge. Therefore, in this paper we will introduce a new way to represent the state.

\paragraph{Action}The action we take in the RL framework is the perturbation we apply to the dynamic graph sequence. Each time we perform an action $a$, we add one edge $(u,v), u,v \in V$ and delete one edge $(u',v'), u',v' \in V$ for all the $\hat{A}_{t} \in \hat{\mathcal{A}}$ in the sequence and obtain a new matrix sequence $\hat{\mathcal{A}}'$:
\begin{equation}
    \hat{A}_{t}' = \hat{A}_{t}([u][v]=1,[u'][v']=0), t\in [1,T] 
\end{equation}

\paragraph{Environment and reward} In our black-box attack on LPDG, the environment is the LPDG model $M$ and the original dynamic graph sequence $\mathcal{G}$. At the attack step $k$, we perform an attack action $a_{k}$ to the edge sequence $\hat{\mathcal{A}}_{k}$ and get the next attacked sequence $\hat{\mathcal{A}}_{k+1}$. We feed it into the model $M$ and get the output ${\hat{E}_{T+1,k+1}^{p}}$. The F1 score metric $f_{k+1}$ is computed by comparing it with the original $E_{T+1}\in G_{T+1}$. Depending on how much the prediction performance decreases from the previous prediction metric $f_{k}$, we give the reward for the action $a_{k}$ and the state $S_{k} = \hat{\mathcal{A}}_{k}$:
\begin{equation}\label{oldreward}
r(S_{k},a_{k}) = f^{k+1} - f^{k}
\end{equation}

\paragraph{Practicable constraints}There are two constraints for the agent to satisfy the practicability requirement. First, for each attack instance $M$ and $\mathcal{G}$, it is only able to perturb a few amount of edges. We set the $\delta$ as the ratio limit and $n$ as the amount limit. For an attack attempt, we could only have $K$ perturbations in total, which satisfies the constraint:
\begin{equation}
  K=min(\delta |E_{max}|,n)
\end{equation}
$|E_{max}| = |V|^2/2$ is the maximum number of edges on the graph and $n$ is a relatively small constant. In the second constraint, the agent only has limited chances to interact with the target model $M$. To learn attacks for one instance, we could only query $M$ within $I$ times during the training.

\subsection{Graph Sequential Embedding}
\label{GSE}
As mentioned before, the state space of the raw matrix sequence $\hat{\mathcal{A}}$ is too large, and makes the deep reinforcement learning difficult to converge. Therefore, we develop a graph sequential embedding (GSE) method, to embed it into a smaller state space. Our GSE method consists of two modules, a static degree embedding  module and a dynamic sequential embedding module.

\subsubsection{Static Degree Feature}

In the static degree embedding, we first compute a degree feature to embed each graph matrix in the sequence first. In step a, we multiply the average adjacency matrix $\Tilde{A} = average(\mathcal{A})$ of the original dynamic graph sequence, with an all-one tensor in the shape of $|V| \times 1$, which gives the average degree of each node in the original sequence. We multiply this again by the average adjacency matrix and get the amount of nodes connected within two steps. This doesn't work exactly for the three steps and the four steps, but we use the same method to  approximate them. We describe this as:
\begin{equation}
d = \{ d_{0},\Tilde{A}d_{0},\Tilde{A}^2d_{0},\Tilde{A}^3d_{0} \}, d_{0}=[1]_{|V|\times 1} 
\end{equation}
In the step b, we normalize each of the last three separately with the batchnorm function $bn$. 
We then {additionally inject} a $|V|\times 4$ random noise feature {by concatenation to enhance the feature's expressiveness~\cite{sato2021random}. Our final degree feature $F$} is represented by:
\begin{equation}
    F = [[Random]_{|V|*4},d_{0},bn(d_{1}),bn(d_{2}),bn(d_{3})]
\end{equation}
This feature is calculated at the start of the attack for each instance, and it remains static throughout the learning.

\begin{figure}[t]
  \centering
   \includegraphics[width=\linewidth]{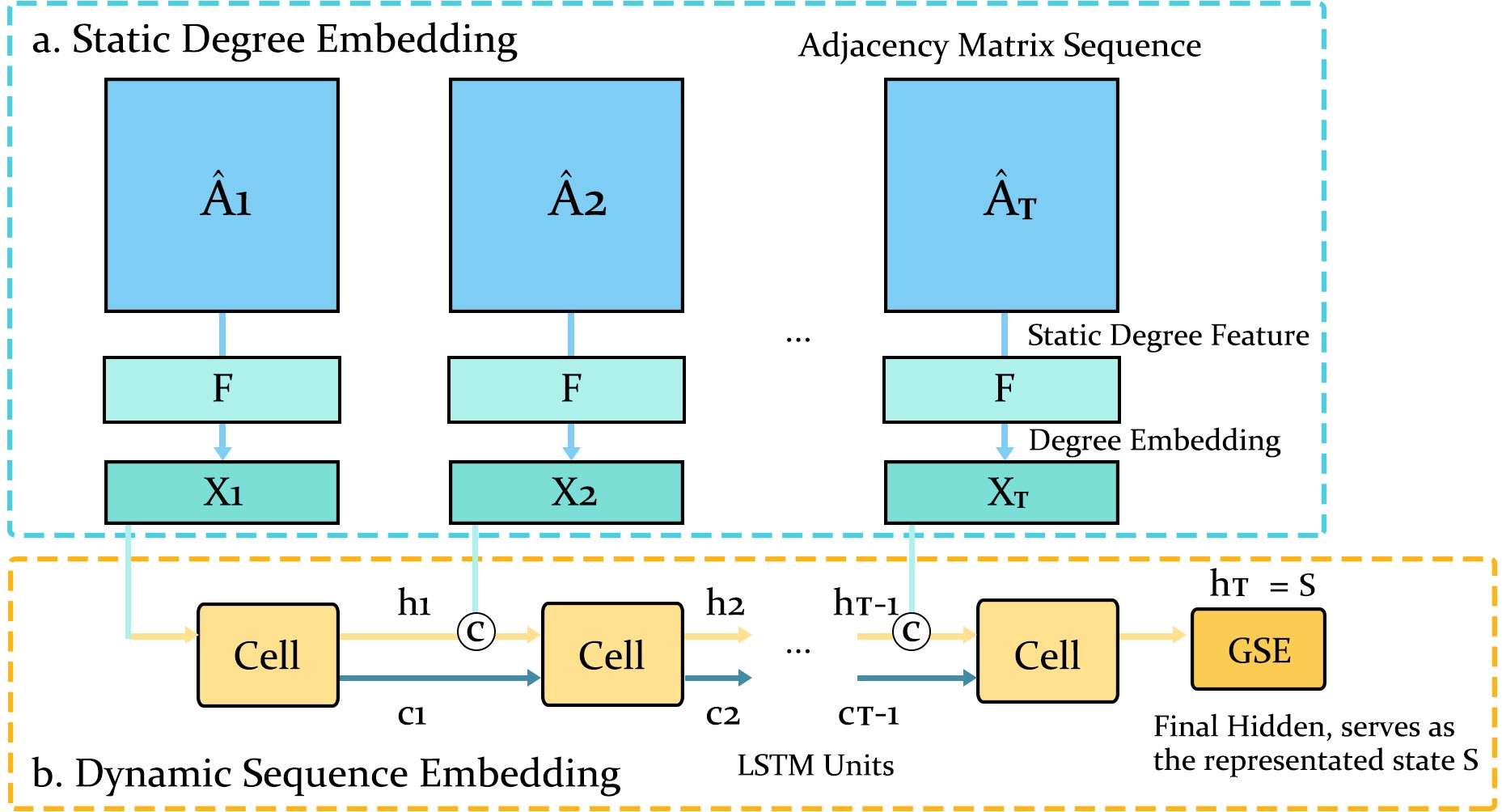}
   \caption{ {Illustration of the proposed GSE method.}}
   \label{fig:GSE}
\end{figure}

\begin{figure*}[!t]
  \centering
   \includegraphics[width=0.9\linewidth]{ 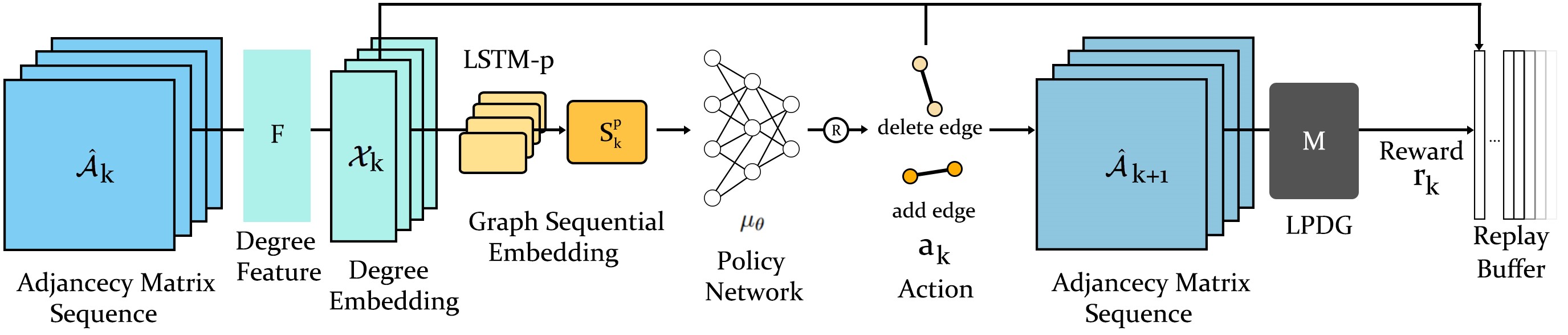}
   \caption{An overview of the agent interaction.}
   \label{fig:Interaction}
\end{figure*}

\subsubsection{Dynamic Sequence Embedding}
After we compute the static degree feature $F$, we multiply it with each matrix $\hat{A}_{t},t\in[1,T]$ to get the degree embedding sequence $\mathcal{X} = \{X_{1},X_2,...,X_{T}\} \in R^{T\times|V|\times|8|}$: 
\begin{equation}
    \mathcal{X} = \hat{\mathcal{A}}F
\end{equation}

This process is described as step a in Figure \ref{fig:GSE}. Then in step b, we use an LSTM module, to dynamically find the efficient
 embedding state $S$ for the graph sequence through training. For simplicity, we denote this process as $SE$:
\begin{equation}\label{eqn:1}\small
\begin{aligned}
h_{0} = X_{1}, & \quad c_{0} = 0, \\
 L_{i}(\_)=W_{i,\_}X_{t}+b_{i,\_},& \quad
 L_{h}(\_)=W_{h,\_}h_{t-1}+b_{h,\_}, \\
 i_t = \sigma(L_{i}(i)+L_{h}(i)) ,&\quad
 f_t = \sigma(L_{i}(f)+L_{h}(f)), \\
 g_t=tanh(L_{i}(g)+L_{h}(g)),&\quad
 o_t = \sigma(L_{i}(o)+L_{h}(o)) ,\\
 c_{t} = f_{t}c_{(t-1)}+i_{t}g_{t}, & \quad
 h_{t}=o_{t}tanh(c_{t}) \\
& \quad \quad \quad \quad \forall t \in [1,T] \\
S = SE&(\mathcal{X}) = h_{T}
\end{aligned}
\end{equation}
In our attack design, we actually use two different GSE models to train embedding states for the actor and the critic. They share the same degree embedding $\mathcal{X}$, but separate LSTM modules and result states. We denote the LSTM modules as LSTM-p and LSTM-q, the embedding processes as $SE_{p}$ and $SE_{q}$, the embedding states as $S^{p}$ and $S^{q}$ for the policy network and the Q network respectively.

\subsection{Multi-Environment Training}
\label{ME}
Our multi-environment training is designed based on a DDPG method. In traditional reinforcement learning, an agent is trained in one or more but identical environments. However, in a practicable attack, we have a limit on the amount of interactions for each attack instance. Hence, we train one agent under several separate instances instead. We attack them alternately and learn our agent from the collective experience they share. These instances are under the same dataset and the same target model, and we believe that they have similarities in the state space and help to explore more efficiently. 

There are three main tasks in our multi-environment training pipeline, (1) perform attack interactions with each instance against the target model; (2) train the Q network and the Q GSE module; (3) train the policy network and the policy GSE module. In each attack step, we perform the three tasks sequentially.

\subsubsection{Attack Interaction}

The Figure \ref{fig:Interaction} describes how our agent performs the attack on each instance. In each attack step $k \in [1,K]$, we have the adjacency matrix sequence $\hat{\mathcal{A}}_{k}$ generated from the $k-1$ step, and the static degree feature $F$ computed in advance. In step a, we apply our graph sequential embedding method, through the degree feature $F$ and the LSTM-p module, and obtain the degree embedding $\mathcal{X}_{k}$ and the graph sequential embedding $S^p_{k}$. In step b, we design our policy network $\mu_{\theta}$ based on an MLP network, denoted $MLP_{\theta}$, to select our add and delete action:
\begin{equation}
    \mu_{\theta}(S^{p}_{k}) = Sigmod(MLP_{\theta}(S^p_{k}))*|V|
\end{equation}

However, there are two situations where we will use a random action instead. First, the interaction records in the shared buffer are smaller than the batch size, so we need to fill it with random attacks first and allow our agent to start training in the following sections. Otherwise, the buffer would be filled with similar actions generated by the untrained policy, leading the agent to a meaningless training result. The second case is that we hope to explore the state space during the attack, so when we go for a certain attack steps we will apply a random action instead. Finally our action $a_{k}$ is selected from:
\begin{align}
\begin{split}
a_{k}= \left \{
\begin{array}{ll}
    \mu_{\theta}(S_{k}^{p}),                    & attack\\
    a_{rand},     & random
\end{array}
\right.
\end{split}
\end{align}
After obtaining the action $a_{k}$, we apply it to the previous sequence $\hat{\mathcal{A}}_{k}$, generate a new graph sequence $\hat{\mathcal{A}}_{k+1}$, perform the prediction on the target LPDG model $M$ and obtain the reward $r_{k}$ in step c. We write the embedding sequence $\mathcal{X}_{k}$, the action $a_{k}$ and the reward $r_{k}$ to the shared replay buffer in step d.


\subsubsection{Q Training}

\begin{figure}[!t]
  \centering
   \includegraphics[width=\linewidth]{ 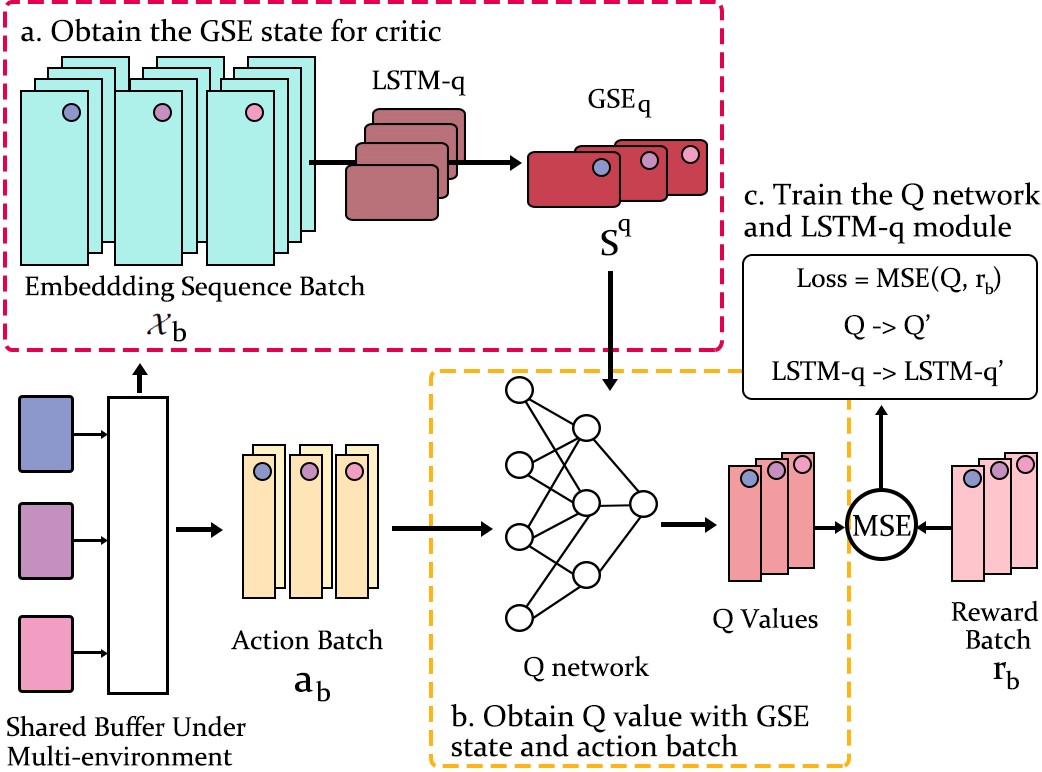}

   \caption{ {Training pipeline for the Q network and Q GSE.}}
   \label{fig:Q}
\end{figure}

The Figure \ref{fig:Q} describes the process of training the Q network and the Q GSE model. The shared buffer has received experience from several instances, and we sample them collectively. The sample consists of a batch of embedding sequences $\mathcal{X}_{b}$, actions ${a}_{b}$ and rewards ${r}_{b} = r(\mathcal{X}_{b}, a_{b})$ from the shared buffer. In step a, we feed them to the LSTM-q to get their sequential embedding states $S^{q}$ for the critic:
\begin{equation}
    S^{q} = SE_{q}(\mathcal{X}_{b})
\end{equation}

In step b, we use an MLP model, denoted as $MLP_{q}$, as our Q network:
\begin{equation}
    {Q}(S^{q},a_{b}) = MLP_{q}([a_{b},S^{q}])
\end{equation}

In the deep reinforcement learning, the value function $Q(S,a)$ is usually defined as: 
\begin{equation}
    Q(S, a) = r(S, a) + \gamma{Q({S'},\mu_{\theta}({S'}))}
\end{equation}
$S'$ is the next state after $S$ takes the action $a$. For the LPDG attack, however, the order of actions is not critical. If a set of actions $[a_1,a_2,..,a_K]$ leads the state from ${S}_{1}$ to ${S}_{K+1}$ and the order is changed arbitrarily, the final state will still be ${S}_{K+1}$. Therefore, we concentrate on learning the Q-value for the reward itself: 
\begin{equation}
    Q({S}, a) = r({S}, a)
\end{equation}
and in step c we use the MSE loss to define the Q loss for the Q network and the Q GSE to train:
\begin{equation}
    loss_{Q} = MSE(Q({S}^{q}, a_{b}), r_{b})
\end{equation}
Finally, we back propagate the Q loss to update the Q network and the Q GSE model with the learning rate $\tau$:
\begin{equation}
Q' \xleftarrow[]{} \tau Q+(1-\tau)Q'
\end{equation}
\begin{equation}
LSTM_{q}' \xleftarrow[]{} \tau LSTM_{q}+(1-\tau)LSTM_{q}'
\end{equation}

\subsubsection{{Policy Training}}

\begin{figure}[!t]
  \centering
   \includegraphics[width=\linewidth]{ 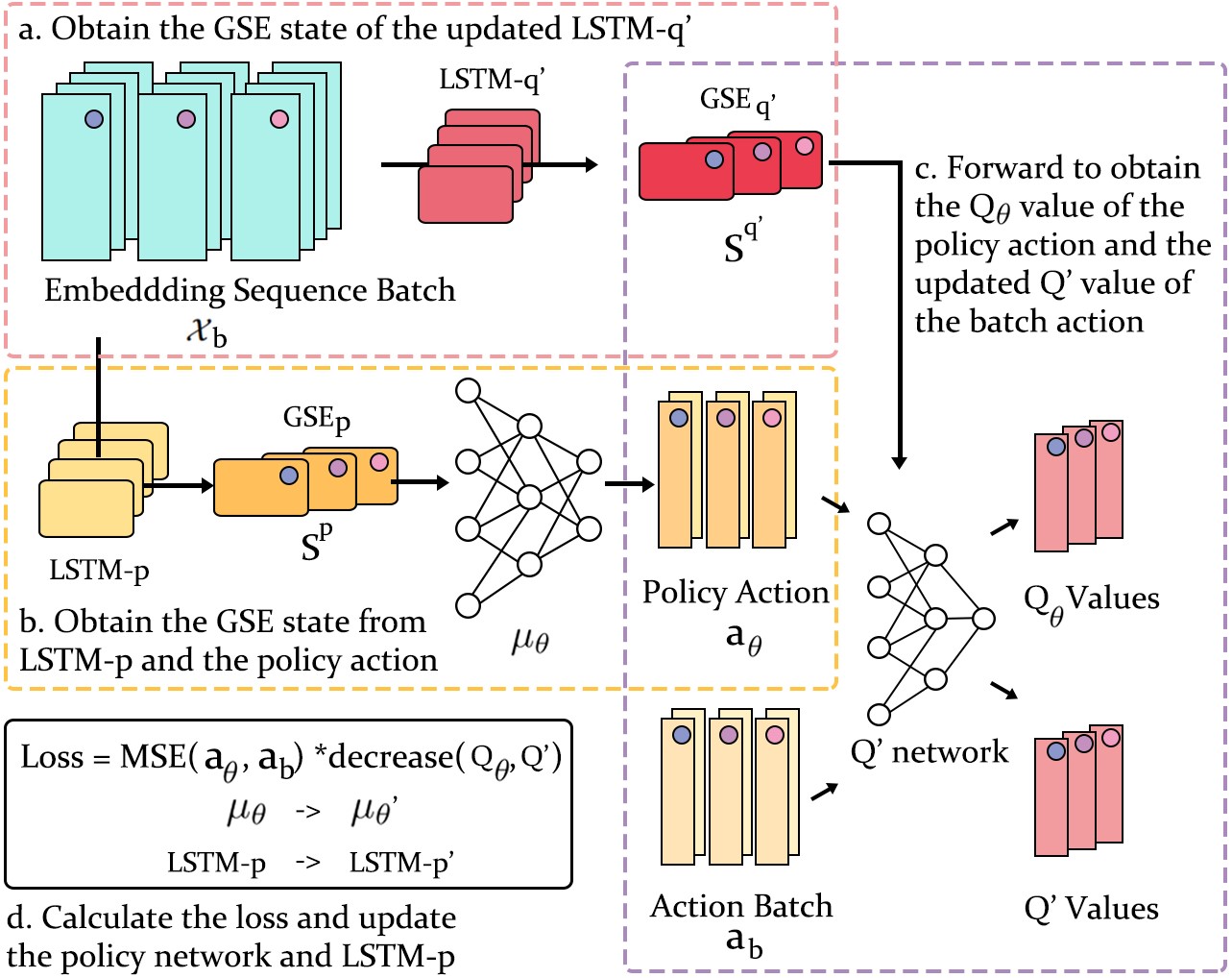}
   \caption{ {Training pipeline for the policy network and the policy GSE.}}
   \label{fig:policy}
\end{figure}
The Figure \ref{fig:policy} describes the process of training the policy network and the policy GSE under the multi-environment. Similar to the Q training, we have the same batch of interaction samples from the shared buffer: embedding sequences $\mathcal{X}_{b}$ and actions ${a}_{b}$. In step a, we forward embedding sequences to the updated LSTM-q' module to obtain the GSE states ${S^{q}}'$ of Q:
\begin{equation}
    {S^{q}}' = SE_{q'}(\mathcal{X}_{b})
\end{equation}
In step b, we forward embedding sequences to the LSTM-p to obtain the GSE states $S^{p}$. Then we feed them to the policy network $\mu_{\theta}$ to get the policy actions $a_{\theta}$:
\begin{equation}
    S^{p} = SE_{p}(\mathcal{X}_{b}), \quad
    a_{\theta} = \mu_{\theta}({S}^{p})
\end{equation}
We define our policy loss as the scalar product of the differences between the actions and the decreases in the Q values. This means that if an action in $a_{\theta}$ is worse than that in the $a_{b}$, we hope to learn towards the batch one, otherwise we hope to stay the same. We update the policy network and the policy GSE with the learning rate $\tau$ as follows:

\begin{equation}
    d({S^{q}}',a_{b}) = max(0, Q'({S^{q}}',a_{b})-Q_{\theta}({S^{q}}',a_{\theta}))
\end{equation}
\begin{equation}
loss_{\mu_{\theta}}({S^{q}}',a_{b}) = MSE(a_{b},a_{\theta}) \cdot d({S^{q}}',a_{b})
\end{equation}
\begin{equation}
\mu_{\theta}' \xleftarrow[]{} \tau \mu_{\theta}+(1-\tau)\mu_{\theta}'
\end{equation}
\begin{equation}
LSTM_{p}' \xleftarrow[]{} \tau LSTM_{p}+(1-\tau)LSTM_{p}'
\end{equation}

\section{Experiments}
\label{sec:Exp}
In this section, we test our attack method on three real-world datasets against  {three LPDG models}, compared with four baseline black-box evasion methods. The experiments consist of two aspects, the performance evaluation and the interaction scale impact test  {for ablation study}. 

For each setting, it consists of the attack method $\mathcal{C}$, the dataset $\mathcal{D}$, the LPDG model $M$, the perturbation constraint $K = min(\delta |E_{max}|,n)$ and the interaction constraint $I$. We first take 10 different instances from the dataset $\mathcal{D}$, each consisting of 11 temporal graphs. The first 10 graphs are the input to  $M$
and the last 1 is the ground truth for the prediction. We use these instances to train the target model $M$, and apply the attack method $\mathcal{C}$ to perform a black-box attack. Each attack consists of several attempts. In each attempt, the agent gives at most $K$ perturbations to the sequence, and the lowest prediction metric achieved during the perturbations is taken as the result of that attempt. The total number of interactions with the target model, taken by all attempts, could not exceed $I$. Once interactions are exhausted, we take the best result of all attempts as the performance of the setting. 

\subsection{Experiments Setting}

\subsubsection{Datasets} 
We use three real-world datasets with varying scales, and   
their properties are shown in Table \ref{datasets}. 
\begin{table}
\begin{center}
\begin{tabular}{|l|c|c|c|}
\hline
Name & Node&Average Edges&Graph type\\
\hline\hline
Haggle & 274 &12584 &Human contact\\
Facebook & 1000 &97779&Social circle\\
AS &6474&141845&Traffic flows\\
\hline
\end{tabular}
\end{center}
\caption{A brief description of datasets.}
\label{datasets}
\end{table}

\paragraph{Haggle} This is a social network available at KONECT and published in~\cite{10.1145/2487788.2488173}, representing the connection between users measured by wireless devices. 

\paragraph{Facebook}This is a subgraph of the ``Social circles: Facebook" social networks from SNAP \cite{leskovec2016snap}. 
We randomly delete edges to generate the dynamic graph sequences.

\paragraph{Autonomous systems} ``AS-733" is a large traffic flow networks available on SNAP~\cite{leskovec2016snap}.
We randomly delete edges to generate the dynamic graph sequences as well.

\subsubsection{Compared Attack}

\paragraph{Random attack}
In this attack method, the agent randomly chooses two nodes to add an edge and two nodes to delete an edge as the action. 
\paragraph{SAC attack}
This attack is introduced by~\cite{fan2021reinforcement}, which claims to be the first black-box evasion attack on LPDG problem. However, without any constraint on interactions, we conclude that this method is impracticable, and prove this in our experiments.
\paragraph {SAC-METP attack}
 {This attack is an ablation version of our method. It applies the multi-environment training pipeline, but takes the same graph embedding method as ~\cite{fan2021reinforcement}.}

\paragraph{GSE attack}
This attack is an ablation version of our method. It applies the graph sequential embedding method, but the models for different instances are trained and applied separately.

\paragraph{GSE-METP attack}
This attack is the complete version of our method. It applies both the graph sequential embedding method and the multi-environment training pipeline.

\subsubsection{LPDG Model}
\paragraph{DyGCN}
Dynamic Graph Convolutional Network (DyGCN)~\cite{dygcn}, is an extension of GCN-based methods. It generalizes the embedding propagation scheme of GCN to the dynamic setting in an efficient manner, and propagates the change along the graph to update node embedding. 

\paragraph{ASTGCN}
Attention Based Spatial-Temporal Graph Convolutional Networks (ASTGCN)~\cite{Guo_Lin_Feng_Song_Wan_2019}, has several independent components. Each of them consists of two parts, the spatial-temporal attention mechanism and the spatial-temporal convolution. In our adaption for the experiments, we only use one component to consist a test ASTGCN.

 {\paragraph{HTGN}
Hyperbolic Temporal Graph Network (HTGN)~\cite{9999499} is a temporal graph embedding methods, which learns topological dependencies and implicitly hierarchical organization of each graph sequence instance individually, and gives link predictions on it.  }

\subsubsection{Attack Settings}
\label{setting}
For the performance evaluation, we set the default attack rate limit to $\delta$ = 0.02, the default attack amount limit to $n = 1000$, and the default interaction limit to $I = 5K$ to make the attack attempts complete in tests. Then, in the interaction impact test, we show our attack results in full on different $I \in [K, 10K]$.

\subsection{Experiments Results}
\subsubsection{Effectiveness Evaluation}
\begin{table}[!t]\small
\begin{center}
\begin{tabular}{|l|c|ccc|}
\hline
Method&Model&Haggle&Facebook&AS \\
\hline
Edge Ratio($\delta$)&&2\%&0.2\%&4.8e-5\\
\hline
Edge Amount($n$)&&751&1000&1000\\
\hline
Interaction($I$)&&3755&5000&5000\\
\hline\hline
Original & DyGCN &0.9930&0.9745&0.8990\\
\hline
Random & DyGCN&0.8979&0.9681&0.8862\\
SAC & DyGCN&0.8149 &0.9665 & 0.8899 \\
 {SAC-METP} &  {DyGCN}& {0.8094} & {0.9651 }& {0.8890}  \\
 {GSE(Ours)} & DyGCN& \textbf{0.8043} &\textbf{0.9629}&0.8910\\
GSE-METP(Ours) &DyGCN&0.8118&0.9653&\textbf{0.8573}\\
\hline\hline
Original & ASTGCN &0.9852&0.9862&0.8303 \\
\hline
Random &  {ASTGCN} &0.9752&0.9298&0.8276 \\
SAC &  ASTGCN &0.9825 &0.9853 & 0.8264\\
 {SAC-METP} &  {ASTGCN}& {0.9750}& {0.9859} & {0.8263}  \\
 {GSE(Ours)} & ASTGCN& 0.9748&0.9407& 0.8226 \\
GSE-METP(Ours) &ASTGCN&\textbf{0.9702}&\textbf{0.9011}& \textbf{0.8020} \\
\hline\hline
 {Original} &  {HTGN} &  {0.9753}& {0.9375} & {0.8665}  \\
\hline
 {Random} &   {HTGN} & {0.8083} & {0.8652} & {0.8377}  \\
 {SAC} &  {HTGN} & {0.9603} & {0.9145}  &  {0.8526}\\
 {SAC-METP} &  {HTGN}&  {0.9676}& {0.9141} & {0.8527}  \\
 {GSE(Ours)} &  {HTGN}&  {0.8311} & {0.8930} &  {0.7944} \\
 {GSE-METP(Ours)} & {HTGN}& {\textbf{0.7417}} & {\textbf{0.8494 }}& {\textbf{0.7649}} \\
\hline
\end{tabular}
\end{center}
\caption{Performance evaluations on the default setting.}
\label{performance}
\end{table}

\begin{figure*}[!t]
\begin{center}
	\begin{minipage}[t]{0.88\linewidth}
        \begin{center}
    
		\includegraphics[width=0.32\linewidth]{ 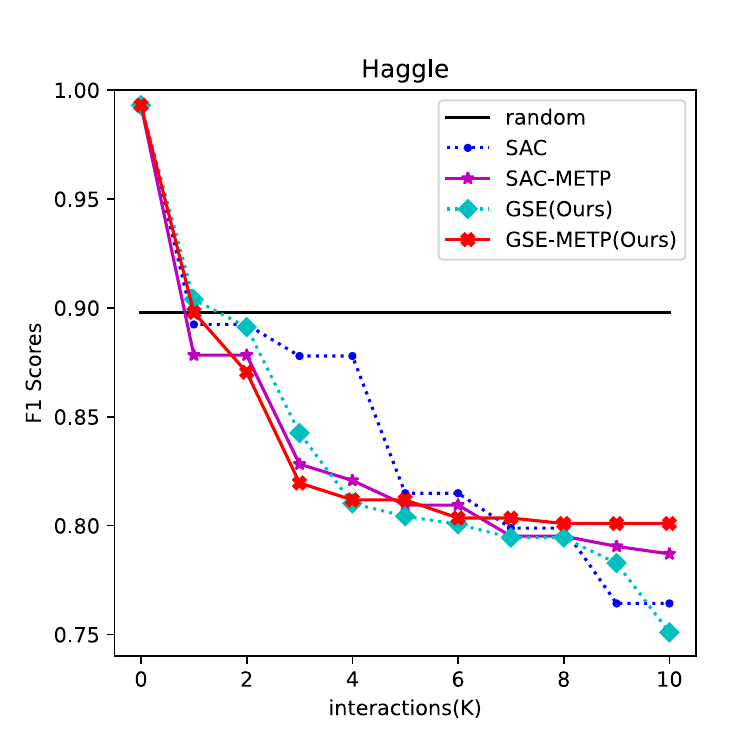}	
		\includegraphics[width=0.32\linewidth]{ 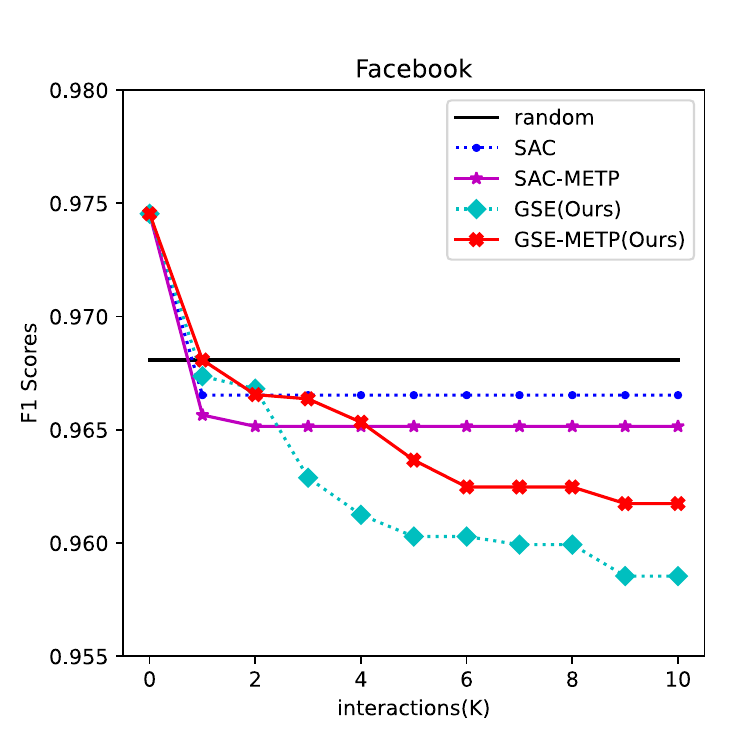}	
		\includegraphics[width=0.32\linewidth]{ 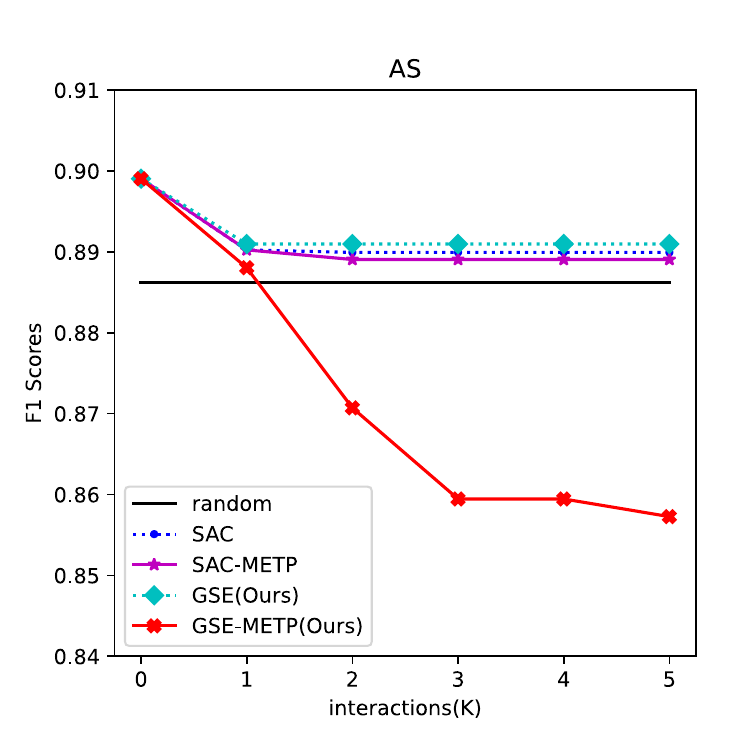}	
        \end{center}
        \begin{center}\footnotesize
        \title{ {(a) DyGCN}}
        \end{center}
	\end{minipage}

	\begin{minipage}[t]{0.88\linewidth}
        \begin{center}
    
		\includegraphics[width=0.32\linewidth]{ 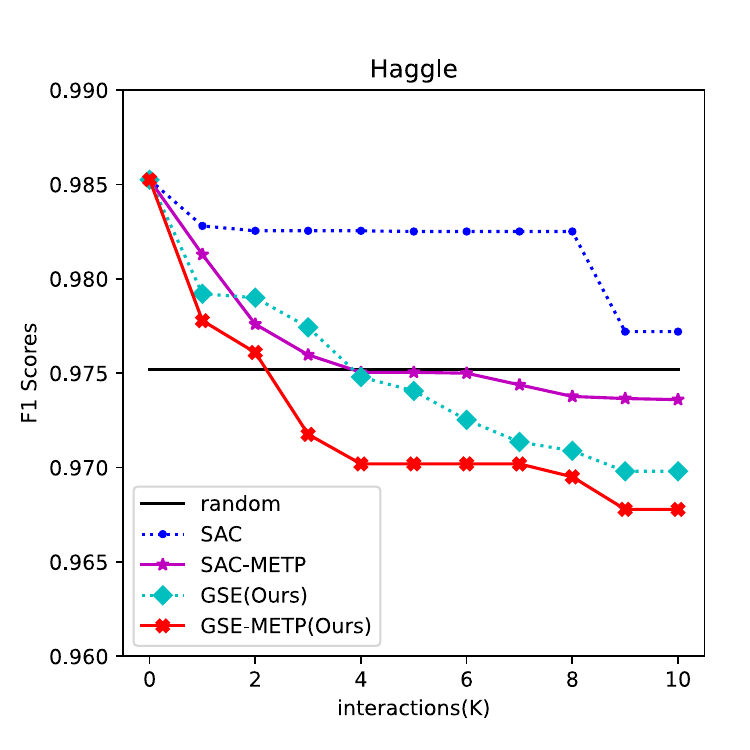}	
		\includegraphics[width=0.32\linewidth]{ 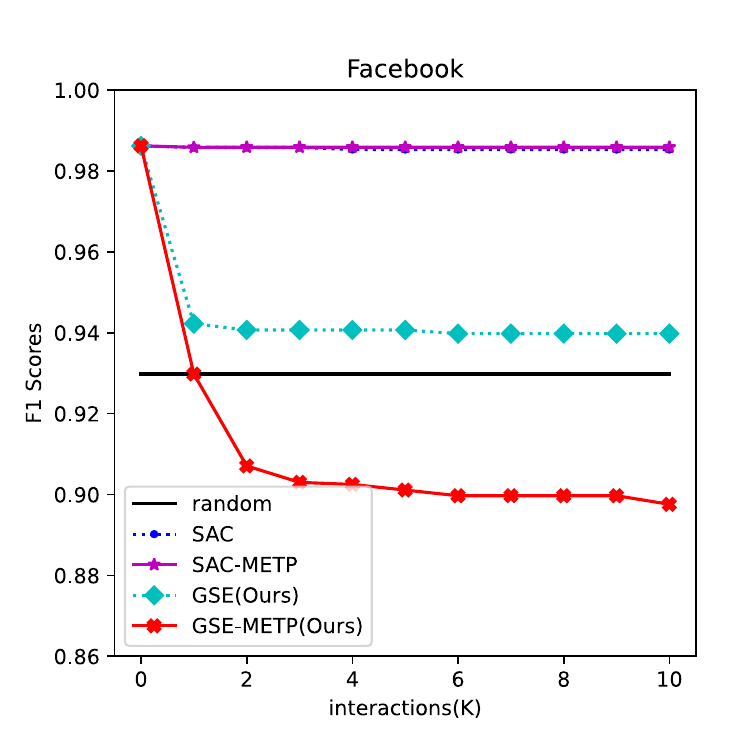}	
		\includegraphics[width=0.32\linewidth]{ 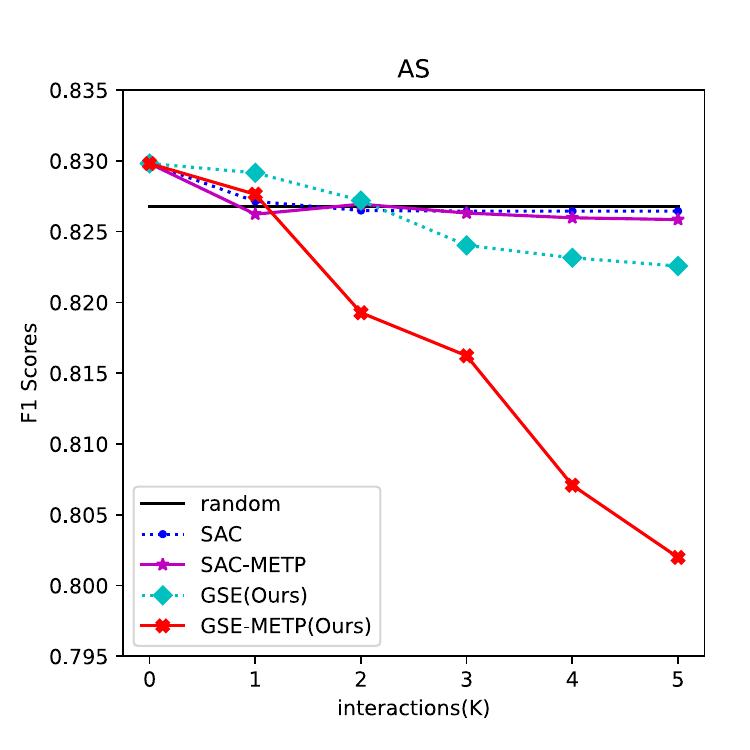}	
        \end{center}
        \begin{center}\footnotesize
        \title{ {(b) ASTGCN}}
        \end{center}
	\end{minipage}
 
	\begin{minipage}[t]{0.88\linewidth}
        \begin{center}
    
		\includegraphics[width=0.32\linewidth]{ 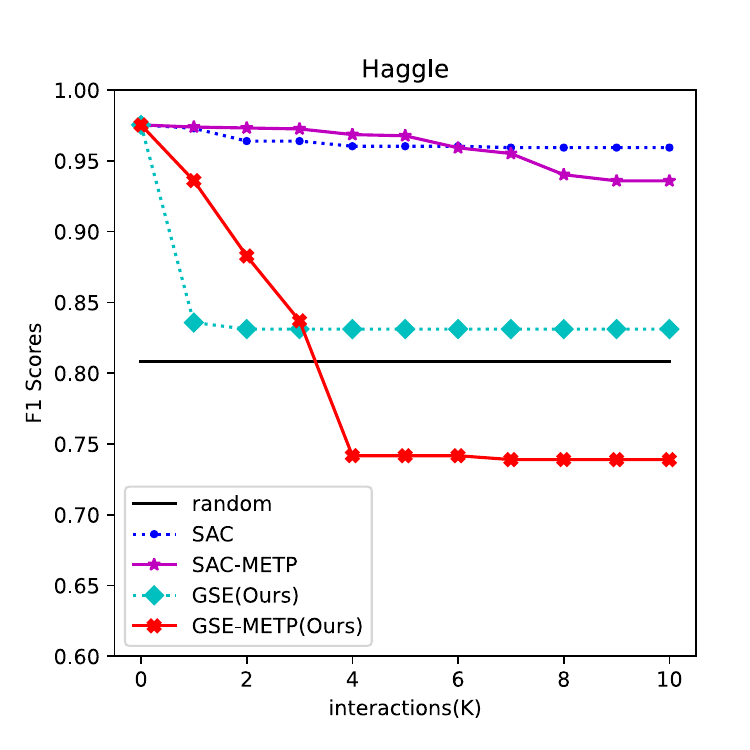}	
		\includegraphics[width=0.32\linewidth]{ 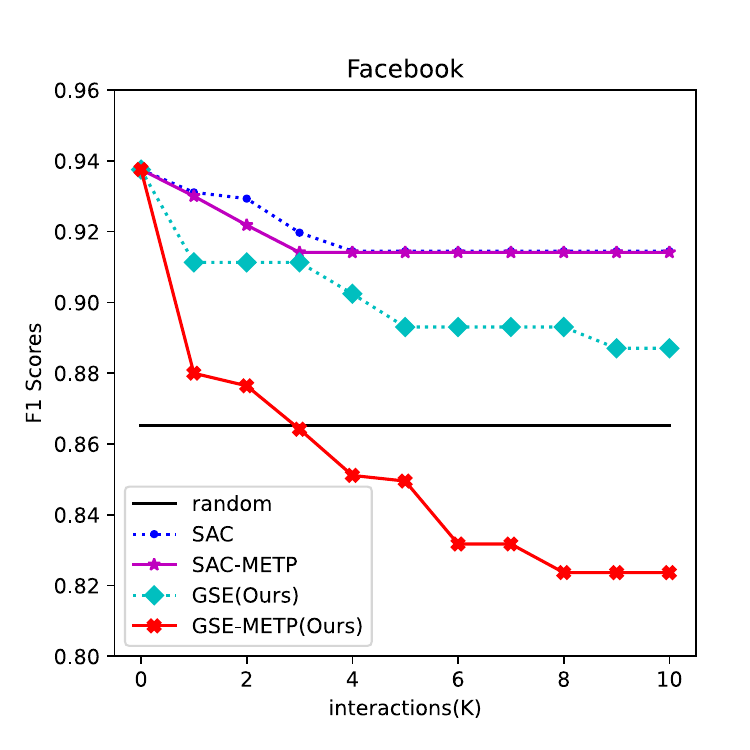}	
		\includegraphics[width=0.32\linewidth]{ 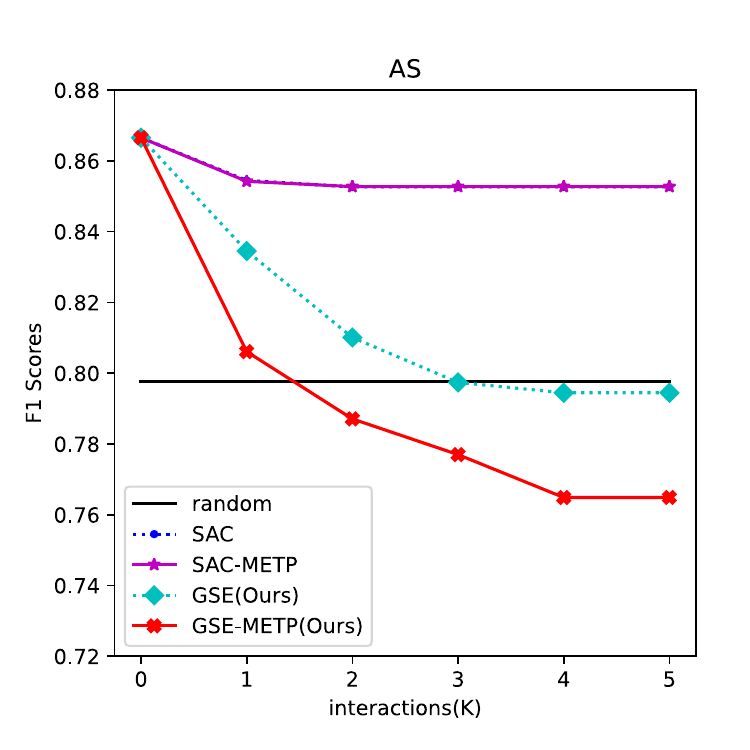}	
        \end{center}
        \begin{center}\footnotesize
        \title{ {(c) HTGN}}
        \end{center}
	\end{minipage}
	\caption{  {Interaction impacts on (a) DyGCN, (b) ASTGCN and (c) HTGN.}}
	\label{fig:Impact}
\end{center}
\end{figure*}
Table \ref{performance} shows the performance of the four attack methods under the default setting against three LPDG models. As shown in the table, our GSE and GSE-METP methods perform best in these tests. Under constraints, the SAC method has good results on the small dataset, while in most cases it behaves close to or worse than the random method. This suggests that it is not a practicable method. Our GSE-METP method behaves effectively on all experiment settings. On the largest dataset AS, GSE-METP has significantly better performance than others. This validates that GSE-METP is the first practicable black-box evasion attack against LPDG methods.

  \subsubsection{Ablation Study} {Figure \ref{fig:Impact} shows our method is more effective as the interaction limit grows compared with other methods. First, on the small Haggle, SAC and SAC-METP are effective when the interactions are enough for training, but fail to converge to good results on the large Facebook and AS. Instead, our GSE and GSE-METP converge to better attacks. This ablation study proves the effectiveness of our GSE design.
}{Second, SAC-METP  and GSE-METP converge faster than SAC and GSE, and also result in better performances. This ablation study shows the efficiency of our multi-environment training pipeline design.}

{\subsubsection{Why SAC fails?}

We further tracked the states and actions during SAC's attack and ours to explore the reason that SAC fails. We found that SAC's represented state has a relatively small variance during the attack, meaning it nearly does not change, while ours changed apparently. 
For instance, on a sample from Haggle on DyGCN, we observed SAC’s add actions converge to the \{0,0\} action, and the variance of its delete actions also decreases from \{8032, 4890\} to \{39, 84\}. In contrast, GSE-METP has average actions on \{97,126\} for adding and \{152, 130\} for deleting and exhibits greater variance \{4655, 4708\} for adding and \{4388, 7405\} for deleting. This suggests GSE-METP adapts its action nodes more responsively in different states, aligning with the ideal agent behavior. As for final performance, SAC only reduces accuracy to 0.854, while GSE-METP to 0.724. 
} 

Based on the observation above, we made an inference for SAC's poor performance: SAC relies on a ranking of node degrees as the state, which is relatively static during learning as each action pair changes the degrees of at most four nodes. Consequently, the generated state representations of dynamic graphs stay nearly constant during the attack process. When the replay buffer is filled with repetitive states, each with a high negative reward, the policy network tends to produce extreme actions to avoid further negative rewards. This results in SAC repeatedly selecting delete-add actions on the same node pair—\{0,0\} or \{V, V\}, which is apparently an undesired behavior.


\section{Conclusion}
We propose the first practicable black-box evasion attack against the link prediction in dynamic graph models. We design a graph sequential embedding method and a multi-environment training pipeline, and combine them with a deep reinforcement learning method, DDPG, to perform effective attacks under interaction and perturbation constraints. Experiments 
on {three advanced LPDG methods} demonstrate the effectiveness of our attack.  
Crucial future work is to design provably robust LPDG 
against the proposed evasion attacks, inspired by existing certified defense on static graphs \cite{wang2021certified,yanggnncert}. 

\section*{Acknowledgments}
Pang is supported in part by Natural Science Foundation of China (62466036),  
Natural Science Foundation of Jiangxi Province (20232BAB212025), High-level and Urgently Needed Overseas Talent Programs of Jiangxi Province (20232BCJ25024) and Key Research and Development Program of Jiangxi Province (20243BBG71035). 
Wang is supported in part by the Cisco Research Award and 
the National Science Foundation under grant Nos. ECCS-2216926, CCF-2331302, CNS-2241713 and CNS-2339686. 

\bibliography{aaai25}

\end{document}